\documentstyle[12pt,aaspp,flushrt]{article}
\singlespace

\def\ang{\thinspace{\rm \AA}}
\def\approxlt{\lower.2em\hbox{$\buildrel < \over \sim$}}
\def\approxgt{\lower.2em\hbox{$\buildrel > \over \sim$}}
\def\bmi{{$(B-I)$}}
\def\eg{{\it e.g.}}
\def\ergf{{ergs cm$^{-2}$ s$^{-1}$ \AA$^{-1}$}}
\def\etal{{\it et~al.\/}}
\def\fnu{{$f_{\nu}$}}
\def\hii{{\rm H}\thinspace{\sc{ii}}}
\def\imk{{$(I-K)$}}
\def\bmk{{$(B-K)$}}
\def\la{\ifmmode {{\rm Ly}\alpha}
        \else {Ly$\alpha$}\fi}
\def\oii{[O\thinspace II]}
\def\p.{^{\prime\prime}\kern-2.1mm .\kern+.6mm}
\def\ten#1{\ifmmode 10^{#1}
    \else $10^{#1}$\fi}
\slugcomment{\it To be published in {\it Astronomical Journal}, Oct 1995}

\begin{document}

\title{Faintest Galaxy Morphologies from {\boldmath $HST$\/} WFPC2 \\
  Imaging of the Hawaii Survey Fields\footnote{Based on observations
  with the NASA/ESA {\it Hubble Space Telescope} obtained at the Space
  Telescope Science Institute, which is operated by AURA, Inc., under
  NASA contract NAS 5-26555.}}
\author{Lennox L. Cowie, Esther M. Hu, and Antoinette Songaila}
\affil{Institute for Astronomy, University of Hawaii, 2680 Woodlawn Dr.,
  Honolulu, HI 96822\\
  cowie@ifa.hawaii.edu, hu@ifa.hawaii.edu, acowie@ifa.hawaii.edu}

\begin{abstract}
We present very deep $HST$ WFPC2 images in the F814W filter of two Hawaii
Survey fields, SSA13 and SSA22.  Using these data with previous ground-based
imaging and spectroscopy, we compare the colors, star-forming properties and
morphologies of the faintest galaxies with a reference sample of bright
nearby galaxies and analyze the changes in field galaxy morphology with
magnitude.  Our principal result is the identification of a new morphological
class of ``chain'' galaxies at the faintest magnitudes.  Based on limited
spectroscopy, we tentatively conclude that these are linearly organized
giant star-forming regions at $z = 0.5-3$ and, if this is correct, that these
are large galaxies in the process of formation.
\end{abstract}


\section{Introduction}

A question fundamental to the understanding of galaxy evolution is how the
properties of galaxy types evolve with redshift, and indeed, whether any
distinctly new classes of objects appear at high redshifts and faint
magnitudes.  Extensive recent work on faint galaxy imaging samples (\eg,
Tyson 1988; Lilly, Cowie, \& Gardner 1991, hereafter LCG; Gardner, Cowie
\& Wainscoat 1993, Glazebrook \etal\ 1994a; Cowie \etal\ 1994, hereafter
Paper I; Cowie \etal\ 1995, hereafter Paper II; Metcalfe \etal\ 1995;
Djorgovski \etal\ 1995; Glazebrook \etal\ 1995a) has explored color and
number-count distributions at both bright and faint magnitudes.  These
surveys find a strong excess of blue objects at faint magnitudes (by
factors of $\sim2$ or more over predictions with no galaxy luminosity
evolution by $B=24$).  Spectroscopic studies (Broadhurst, Ellis, \& Shanks
1988; Colless \etal\ 1990; LCG; Cowie, Songaila, \& Hu 1991, hereafter
CSH;  Loveday \etal\ 1992; Tresse \etal\ 1993; Lilly 1993; Songaila
\etal\ 1994, hereafter Paper III; Glazebrook \etal\ 1995b) have shown that
the bulk of galaxies studied to $K=19$, $I=22$, and $B=24$ have redshifts
$z\approxlt1$ and that a large fraction of the strong excess of very blue
objects at faint magnitudes can be identified with relatively low-mass
gas-rich systems at $z\sim0.4$ (CSH).

In the present paper we study {\it morphological} changes in field galaxies
with redshift using very deep (4.3--8 hr) $HST\/$ WFPC2 images in the
\hbox{`wide-$I$'} (F814W) filter on two Hawaii Survey Fields, SSA13 and
SSA22.  The substantial increase in depth over previous field
investigations with $HST\/$ (\eg, Griffiths \etal\ 1994; Glazebrook \etal\
1995c) permits studies of the very faint-end morphologies and colors, and
the construction of $I$-band number counts down to $I$=26; while the
availability of previous multi-color imaging on these fields (Paper I) and
related wide-field surveys (Paper II), in combination with near-complete
spectroscopic coverage over selected subregions (Paper III), allows
quantification of changes in color and star-forming properties with
redshift. We present a reference atlas of bright nearby galaxies, with
$K\leq14$, spectroscopically studied and with similar linear resolution and
S/N to the $HST\/$ observations, and compare it with the assembled atlas of
objects from the $HST\/$ samples.  In addition, as we have stressed in
Paper I, the $K$-band data on these samples permits a magnitude-limited
sample selection that is not biased by the type mix of galaxies being
studied, since magnitudes selected in the $K$-band, as opposed to the $B$-
and even the $I$-band, are insensitive to highly variable K corrections
that, at optical wavelengths, depend on both the morphological type and
redshift of the objects measured.

The major result of the present paper is the identification of a new
morphological class of objects at the faintest magnitudes.  We call these
`Chain' galaxies, and they appear as extremely narrow, linear structures
(barely resolved by the WFC in the transverse direction) with superposed
bright `knots'.  The chains can be identified with a spectroscopically
distinct class of very blue (flat-spectrum) objects, which onsets in
substantial numbers (about 2 arcmin$^{-2}$ at $B\sim24$) at approximately
$B=23$ (cf.\ LCG), and which comprises 20--50\% of the population identified
with the strong excess in blue number counts.

The acquisition, processing, and calibration of the WFC2 data are described
in \S2.  A summary of the number counts and colors, showing the strong
blueing trend at faint magnitudes is given in \S3.  A qualitative description
of the galaxy morphologies considered in terms of various color selection
criteria is presented in \S4, along with a description of the
characteristics and statistics of the faint blue populations. \S5
discusses the properties of the chain galaxies in the context of possible
models for their formation and their extreme linear structure.  A
quantitative analysis of the galaxy morphologies and surface brightness
profiles will be given in a second paper (Hu \etal\ 1995).

\section{Observations}

WFPC2 integrations were taken on two of the Hawaii Galaxy Survey fields
(which are defined in LCG and Paper I): the SSA22 field was observed with
the F814W filter on the WFPC2 (twelve 2400 s exposures on 1994 November 21
and 26) and the SSA13 field in the same configuration (six 2600 s exposures
on 1995 January 30).  Each exposure was dithered by a five arcsecond step
from the previous one.  The pipeline-processed flattened and calibrated
images were used.  The sky was determined from the median of the individual
frames and a normalized version subtracted from each frame.  The frames were
then resampled to $1600\times 1600$ pixels and registered.  Cosmic rays were
removed by comparing each pixel with the median of the pixels in all the
frames and rejecting those which deviated by more than $3\sigma$.  The
frames were finally summed.  The images of the three WFC2 frames on each
field are shown as Figures 1 and 2 (Plates 1 and 2).  The final full width
half maximum (FWHM) was measured as $0\p.22$ on the SSA22 images and
$0\p.19$ on the SSA13 image.

The photometric system was derived from the PHOTFLAM calibration generated
in the pipeline process with 0 mag corresponding to the Kron-Cousins $I$-band
zeropoint of $1.22\times\ten{-9}$ \ergf\ from Bessell (1979).  For the WFC
chip 2 this corresponds to
\begin{equation}
     I_{HST} = 30.12 - 2.5\,{\rm log_{10}(DN)}
\end{equation}
where DN is the number of counts in a 2400 s interval in the F814W filter.
For the faint galaxy photometry of the present paper we have adopted
corrected aperture magnitudes using a $1\p.6$ diameter aperture offset by
$-0.16$ mag to correct to total magnitudes.  The offset was measured by
comparison with $6''$ diameter aperture magnitudes.  The $1\p.6$ diameter
aperture is the smallest usable aperture in which the correction to total
magnitude is relatively invariant from object to object.  This is
illustrated in Fig.~3, where we show the difference between $1\p.6$ and
$3\p.2$ diameter aperture magnitudes for isolated objects in the fields; the
spread in total magnitudes corresponds to an uncertainty of approximately
$\pm0.2$ mag in the total magnitude of an object, which we consider
acceptable for the present type of work.  For smaller apertures the spread
rises rapidly.

The $I_{HST}$ magnitudes closely approximate the ground-based $I$ magnitudes
of Paper I.  In Fig.~4 we show $(I_{HST}-I)$ vs \bmi\ where $I$ and $B$ are
from papers I and II.  The least squares fit gives a small color term $-0.08
+ 0.05$\bmi\ and a dispersion of $0.14$ mag.  For the present purposes we
ignore the color term and use $I_{HST}$ and $I$ interchangeably.  Because of
the characteristic flat spectrum of many sources, it has sometimes been the
convention to quote object magnitudes in terms of $AB\/$ magnitudes -- a
system where the zero-points for the various color bands $B_{AB}$, $I_{AB}$,
and $K_{AB}\/$ coincide for objects which are flat in \fnu.  Following the
discussion of LCG, we give the conversion between Kron-Cousins $I\/$
magnitude and $I_{AB}\/$ as $I\sim I_{AB}-0.48$, and $B \sim
B_{AB}+0.17$\ and $K \sim K_{AB}-2.0$\ for Johnson $B$ and $K$.  We shall
occasionally move to the $AB$ system where it is appropriate.

For each field a primary catalog was generated of all objects with surface
brightness above 24.7 mag arcsec$^{-2}$ in a $0\p.2$ boxcar-smoothed image.
Objects with centers closer than $1\p.6$ were then combined and the magnitudes
measured following the procedure outlined above.  The star-galaxy separation
was made using the inverse second moment classifier (Kron 1980).  $B$ and
$K$ magnitudes for the $I$-selected samples were obtained for the object
list from the ground-based data using the procedures outlined in Paper I.
In addition, $I$ magnitudes were measured from the $HST\/$ data for
ground-based $K$- and $B$-selected samples.

The noise levels for each field were next determined by measuring the
dispersion of the signal from blank sky positions on each image.  The
$1\sigma$ level for the corrected total magnitudes is similar for the 13-
and 22-hr fields at $I=27$, since the 13-hr field is at much higher
ecliptic latitude and the background is 22 mag arcsec$^{-2}$ vs 21.5 mag
arcsec$^{-2}$ for the 22-hr field.

Finally, the incompleteness of the object recovery was measured as a function
of magnitude.  We adopted the procedure of adding, to the data in each WFC2
chip, field data from another of the two WFC2 fields reduced by a factor of
10, and then re-running the cataloging procedure.  The fraction of the added
objects which were retrieved was then determined.  The incompleteness becomes
substantial at magnitudes greater than $I=25$:  only 52\% of objects between
$I=25$ and 26 were recovered.  We have therefore computed the galaxy counts
only to $I=26$.

\section{Counts and Colors}

The $I$ band counts from the two fields are shown in Fig.~5.  They extend the
previous deepest $I$-band counts from ground based data by about 1.5 mag to
$I=26$ ($I_{AB}=26.5$), and there is good agreement with the ground-based
data in the region of overlap (Tyson 1988; LCG).  The faint $I$ counts are
well fit by the power law $dN/dm = 240\times\ten{0.37(I-17)}$ galaxies
deg$^{-2}$ mag$^{-1}$.

The galaxy counts are now becoming very well defined to extremely faint
magnitudes over the full optical and near-IR range.  Fig.~6 shows a
compilation of recent faint $K$-band counts from a number of groups.  These
are well fit by $dN/dm = 1460\times\ten{0.25(K-15)}$ from $K=17$ to $K=24$,
or equivalently, $K_{AB}=26$.  The faint $B$-band counts have recently been
summarized by Metcalfe \etal\ (1995) and to $B=27$ ($B_{AB}=26.8$) have the
form $1.0\times\ten{0.42(B-14)}$.  The counts in each color are compared in
Fig.~7, which shows just how remarkable the blueing trend in the galaxy
population really is.  At the faintest $AB$ magnitudes ($AB=26$) the $K$
counts have fallen below the $I$ counts, and both have closely approached the
$B$ counts.  This requires that the vast majority of the faintest galaxies
are very blue, with spectra close to flat $f_{\nu}$ over the full range
$4000\to25,000$\ang .

At slightly brighter magnitudes we can investigate the colors of individually
selected galaxies.  Fig.~8 shows the \imk\ colors of a $K=22$ sample selected
from the ground-based data.  The substantial improvement in the S/N of the
$I$-band data obtained with $HST\/$ now allows us to see that at the faint
end the population is dominated by blue \imk\ galaxies, with a small number
of very red \imk\ objects.  At $K=21\to22$ the median color is therefore very
blue with $\langle$\imk$\rangle=2.1$\ (only 0.6~mag redder than flat
$f_{\nu}$), but this is determined by  80\% of the objects (12 out of 15).
The remaining 20\% (3 out of 15) are quite red.

Fig.~9 shows the \bmi\ colors of an $I\leq25$ sample selected from the full
$HST\/$ data.  At $I=25$ the median \bmi\ color of a galaxy is only 1.2,
less than 0.5 mag different from flat-spectrum.

\section{Galaxy Morphology}

\subsection{$K$-Selected Samples}

For the purposes of studying the morphology of normal galaxies at $z\sim1$
it is almost essential to use a near-infrared selection, since even an
$I$-band selection results in K correction biases against higher redshift
elliptical and spiral galaxies (Paper I).  As a result, direct selection from
the $HST\/$ data results in a serious bias to lower $z$ and blue galaxies.
For our primary analysis, therefore, we have used the $K$-band selected samples
of Papers I and III to define the sample, and then used the $HST$ $I$- band
images to study the morphology.

At $K\leq18.5$ (Fig.~10, Plate 3) we show all the objects in the $HST\/$
fields.  The $K$-band magnitudes (lower left corner of each panel) are taken
from the $K$-band strip data of Paper III.  In order to provide uniform
contrast we have normalized each image by its $I$-band flux so that sky
noise levels appear higher for the fainter objects.  All of the objects have
spectroscopic identifications either from Paper III or from subsequent
observations with the LRIS spectrograph on the Keck 10 m telescope (Songaila
\etal\ 1995).  The galaxy redshifts are presented in the lower right corner
of each panel.  Fig.~11 (Plate 4) shows the corresponding data for
$K=18.5-19.5$ selected from the strips for the 22-hr field where the strip
image is much deeper, but only in the deep field region of Paper I in the
13-hr field.  Finally, Fig.~12 (Plate 5) shows the data for a $K=19.5-20$
sample selected from only the deep field $K$ samples of Paper I.  All but
three galaxies of the $K<19.5$ sample have measured redshifts, but only
slightly less than half of the $19.5-20$ sample do.

We can see from the plates that at $K\leq19.5$ we appear to be selecting a
rather normal population of galaxies.  The redshift data show that nearly
all of these objects are at $z\approxlt1$.  For comparison we have shown a
ground-based complete $K\leq14$ selected sample (Fig.~13, Plate 6) of 25
spectroscopically identified galaxies displayed at a similar linear scale
and S/N as the $HST\/$ data.  As we shall address more quantitatively in the
second paper (Hu \etal\ 1995), the distribution of types and sizes is
remarkably similar between the bright and faint samples.

At $K\geq19.5$ the morphologies of the galaxies become much more diverse.
Fig.~12 continues to show some galaxies with relatively normal appearance,
but there is now a large fraction of amorphous and distorted galaxies, as
well as a population (\#s 6, 16, and 19) of just-resolved, symmetric objects
which are very faint in the $I$ band.

The onset of the morphology change corresponds to the onset of the blueing of
the $K$-selected samples, and the bifurcation into blue and red populations
in the color-magnitude diagram of Fig.~8; it may also be associated with the
movement of the maximum redshift beyond $z=1$.  At $z>1$, as the 4000\ang\
break moves through the $I$ band, galaxies fade rapidly and the redder
objects in the $19.5-20$ sample most probably correspond to the
high-redshift $(z>1)$ tail of spirals or ellipticals (Paper I).

The simultaneous appearance of the amorphous galaxies is more surprising.
For the small number of galaxies where redshifts have been measured these
are emission-line objects at moderate redshift $(z\sim0.4)$.  At
$K=19.5-20$ they constitute almost half the objects and completely dominate
the blue populations.  The most likely explanation is that because of
volume geometry effects and/or galaxy evolution we are seeing relatively
few high-$z$ galaxies, and that as we move to fainter magnitudes our
samples are rapidly becoming dominated by the faint end of the luminosity
function of modest redshift galaxies.  The faint end slope of the counts
and the blueing trend then correspond to the shape and colors of the faint
end of the luminosity function.

At fainter magnitudes we have constructed a $K=20.5-22$ atlas for the 22-hr
deep field and a $K=20.5-21.5$ atlas for the 13-hr deep field (Figs.~14 and
15, Plates 7 and 8).  There is very little redshift information for this
faint end sample; however, at these $K$ magnitudes the population is
dominated by the blue \imk\ galaxies: only 9 objects have \imk\ colors even
in excess of 3.  Unusual morphologies also dominate at the faint end:
roughly half of these unusual objects are the amorphous galaxies that are the
majority of the $K=20$ counts, but now we also see a considerable population
of a new class of object which appears narrow and linear, and often has a
beaded morphology.  Prototype examples are objects 4, 5, 9, and 15 of
Fig.~14.  A substantial fraction of the $K=20.5-22$ galaxies fall into this
category.  They appear similar to the morphological class of chain galaxies
in the Arp atlas (1966) and we shall refer to them as chain galaxies in the
subsequent text.

To loosely quantify these changes in the mix of morphologies we have adopted
a crude galaxy classification in which galaxies with a strong nucleus and
surrounding smooth envelope are classified as Type 1, those with a strong
nucleus and apparent spiral arm structure as Type 2, and those with an
amorphous appearance as Type 3.  Very roughly, Type 1 includes the Hubble
types E/S0/Sa, Type 2 the Sb's and Sc's, and Type 3 the Sd's and Im's.
We have also distinguished two further classes of galaxies -- Type 4,
corresponding to galaxies with strong signs of interaction, and Type 5,
corresponding to the chain galaxies described above.  Glazebrook \etal\
(1995c) have used a similar breakdown in analyzing the Medium Deep
Survey but with our types 3--5 combined into a single category of peculiar
and merging galaxies.  The distinction between Type 1 and Type 2 is poorest,
and becomes very uncertain at the faintest magnitudes, but the remaining
types are easily picked out.

We have summarized the distribution of types vs $K$ magnitude in Fig.~16.
At $K<19$ nearly all galaxies are Type 1 or Type 2 in roughly equal numbers
with a small admixture of interacting galaxies (Type 4).  At $K>19$ the Type
3 galaxies appear in substantial numbers, while at $K>20.5$ the chain
galaxies of Type 5 become the largest class.

As Glazebrook \etal\ (1995c) have emphasized, this rapid evolution in
morphology has major implications for interpreting galaxy number counts.
In Fig.~17 we show the fraction of galaxies in the $K$-band counts that fall
into the Type 1 and 2  categories.  This `normal' fraction has dropped to less
than 20\% at $K=21$.  Glazebrook \etal\ have argued that the $B$-band galaxy
counts may be modelled by including the normal population (here Types 1 and
2) with no evolution in the number density or luminosity, and then adding the
rapidly evolving population of peculiar and interacting galaxies (here Types
3--5).  However, this model does not provide an adequate description of the
$K$ counts.  As Gardner \etal\ (1993) have shown, there is an excess of $K$
counts at $K=17$ over models in which there is no luminosity evolution.
However, from Figs.~16 and 17 we can see that at this magnitude there is
very little contamination by the Type 3--5 populations which therefore
cannot be the explanation for this excess.

It is probable that the Type 1 and 2 galaxies are also evolving, and this
evolution may well be correlated with the onset of the other types.  If
this is true we will need to develop a wider understanding of the nature of
the evolution before we can begin fully to reinterpret the number counts.

\subsection{Blue-Selected Samples}

Because the anomalous faint-end population is so blue we have also considered
the morphology in the $B$-selected sample.  For each of the two fields a
complete $B=24.5$ atlas was formed in the three $HST\/$ WFC areas.  The 71
objects in the sample are shown in Figs.~18 and 19 (Plates 9 and 10).  Each
frame in these plates is $3\p.2$ on a side. The $B$ magnitude is shown in the
lower left corner and the redshift (where known) in the lower right corner.
At least 6 of the magnitude-selected sample are stars based on the
spectroscopic observations while 21 of the objects are very blue, satisfying
the color criterion \bmi$<1.4$ (Paper I).  These latter objects are marked
`flat' in the upper right hand corner.  All but one of the spectroscopically
observed objects at $B\leq24$ have been identified.  At $B\leq23$ all objects
are at $z\leq1$ whereas for $23\leq B\leq24$, 6 of the 23 objects with
redshifts are at $z>1$.  This redshift distribution in the {\it HST\/} fields
is consistent with the much larger sample of Songaila et al.\ (1995).

Inspection of Plates 9 and 10 shows that the flat spectrum galaxies are split
into several morphological classes.  There are low-redshift extragalactic
\hii\ regions and Im galaxies such as object 8 in Fig.~19 and object~4 in
Fig.~18, which are at very
modest redshifts.  These are small diffuse objects, and roughly a third of the
$B\leq24.5$ flat-spectrum objects fall into this morphology class.
However, many of the remaining flat objects consist of the chain galaxies
noted in the $K$-band data.  A further unusual multicellular morphology
which may be related in some way to the chains is seen in object 26 of
Fig.~18, where a set of blobs is immersed in a more diffuse circular
envelope.  The object with peculiar morphology reported by Glazebrook
\etal\ (1994b) may be a bright example of this class.  The remaining flat
spectrum galaxies have single or slightly elongated very marginally
resolved images.  Examples are objects 8 and 9 in
Fig.~18.  If the chain galaxies are linear, some of these objects could
correspond to chain galaxies seen in projection.

Clearly defined chain galaxies comprise 6 of the 41 galaxies in the
$23.5\to24.5$ magnitude range (objects 11, 23 and 29 in Fig.~19 and objects
15, 17, and 25 in Fig.~18) while more questionable cases could roughly
double this number.  Thus the chains constitute $15-30$\% of the objects at
$B=24$ and do not comprise the majority of the excess blue counts seen at
this magnitude, which appears to be produced by a combination of the chains
and of the excess number of low redshift galaxies.  The exact value of the
blue excess at $B=24$ is uncertain, but lies between a factor 2 and 4 (\eg,
Tyson 1988; LCG; Metcalfe \etal\ 1991; Glazebrook \etal\ 1995a).  This
means that in rough terms the chain galaxies may contribute $20-50$\% of
the excess at this blue magnitude.

\section{The Chain Galaxies}

The chain galaxies seem to be a new class of object which appears in the
faintest magnitude samples and is associated in part with the
extreme blueing seen in the faint galaxies.  Their curious morphologies and
high spatial frequency make them an extremely interesting class.

To determine the spread in the properties of the chain galaxies, we have
formed an atlas of all galaxies with $I < 25$\ in the $HST$\ fields that
clearly show the chain morphology.  The 26 objects selected in this way
from the two fields are shown in Figures 20a--c (Plates 11, 12 and 13) and
their properties are summarized in Table~1 where we give the $K$, $I$\ and
$B$\ magnitudes, the transverse width when the point spread function is
deconvolved, and an axial ratio, defined here as the ratio of the
longitudinal to transverse lengths of the undeconvolved surface brightness
contour corresponding to 20\% of the peak surface brightness.

The chain galaxies are uniformly blue at the brighter magnitudes where the
colors can be measured reliably.  In Fig.~21 we compare the \bmi\ colors of
chains with $I < 24$\ with the full sample of $I < 24$\ galaxies.  The chains
clearly fall along the lower envelope of the field galaxies, having a median
\bmi\ = 1.6.  The uniformly blue color of the sample argues that, irrespective
of the origin of the light, the objects cannot be at redshift $z \gg 3$\ since
otherwise some would lie behind intergalactic clouds which are opaque beyond
the wavelength of the Lyman limit and would extinguish the $B$\ light.  (cf.\
Madau [1995] for a recent discussion of this effect.)  The chains are also
extremely blue
in \bmk\ (Table~1) and are generally only marginally detected in {\it K}.  This
rules out the possibility that we are seeing relatively normal galaxies in the
rest ultraviolet and that the peculiar morphologies are a consequence only of
the distribution of the star-forming regions.

Morphologically, we find that the chains are extremely narrow in their
transverse extent, being only marginally resolved in the transverse
dimension with sizes $\sim0\p.05-0\p.1$.  The longitudinal sizes are around
$2-3''$.  (The measured dimensions
are summarized in Table 1.)  These very large ellipticities argue strongly
against these being any class of edge-on object.  The linear extent of each
blob in the chain is similar to their transverse dimensions while blob
separations are about $0\p.5$, or several times larger (cf.\ Fig.~22).

The extremely narrow transverse widths argue against the chain population
being local, since at redshifts where the Euclidean approximation is valid
the transverse dimensions are only
\begin{equation}
     r= 10{\rm\ pc}\left( {{\displaystyle z}\over{\displaystyle 0.01}}\right)
     \left( {{\displaystyle \sigma}\over{\displaystyle 0\p.05}}\right)
     h_{75}{}^{-1}
\end{equation}
where $h_{75}$ is the Hubble constant in units of 75 km s$^{-1}$ Mpc$^{-1}$
and $\sigma$ is the transverse dimension in arcseconds.  If we assume instead
that the chains are at redshifts $z \approxgt 0.5$ then the individual blob
dimensions correspond to a linear size of $\sim0.5-1\ h_{75}{}^{-1}$ kpc.
The spread in the transverse sizes is also quite small, only a factor of
four.  This also suggests that the objects must lie at $z \approxgt
0.5$\ where the angular distance becomes relatively invariant since otherwise
they would have to be confined to a narrow shell in redshift.

We have some limited spectroscopic information on the brighter chain
galaxies.  Chains~0 and 1 of Fig.~20a are emission line galaxies at $z =
0.489$\ and $z = 0.505$, respectively.  Two further objects, SSA22--16 and
SSA22--24, were identified as unusually blue in the LCG sample and were the
subject of intensive study.  SSA22--16 (which is object 4 in Fig.~14 and
object 3 in Fig.~20a) has a strong emission line and appears to lie at $z =
1.36$\ (Fig.~23).  In contrast, the optical spectrum of SSA22--24 (object 9
in Fig.~14 and object 6 in Fig.~20a)is featureless except for a continuum
break near $4000\ang$\ in the observed frame (LCG).  The simplest explanation
of the color and spectral information is that the chain galaxies are intense
star forming galaxies which, if the \oii\ line has redshifted out of the
optical window at $z>1.6$, appear as relatively featureless flat spectrum
objects until the \la\ wavelength moves into the observable window at
$z\sim2$.  \la\ emission is weak, analogous to low-redshift extragalactic
\hii\ regions and Im's, but the imposed signature of the intergalactic
\la\ forest produces a partial break at shorter wavelengths (Madau 1995).
This interpretation of the break in SSA22--24 would then place it at $z=2.4$.

While recognizing the indirect nature of these arguments and the limited
nature of the spectroscopic information, we therefore suggest that the chain
galaxies consist of star-forming regions and lie in the redshift range, $z
\sim 0.5 - 3$.  At these redshifts the typical chain would contain about
$\ten{9}$\ OB stars and have a mass larger than or comparable to that of a
present-day galaxy.

This brings us finally to the strangest feature of the galaxies --- their
linear structure.  Given the ubiquity of these objects, we assume that this is
not a gravitational lensing phenomenon but rather represents their intrinsic
nature.  Since the linear morphologies are inherently unstable and will disrupt
on a timescale comparable to the transverse crossing time which is only
\begin{equation}
    t = 3\times\ten{7}\ {\rm yr}\ \left({{r} \over {1\ {\rm kpc}}}\right)\
    \left({{10\ {\rm km\ s}^{-1}} \over {v}}\right)
\end{equation}
where $v$\ is the transverse velocity,
it is probable that we are seeing the structures in the process of formation.
The simplest explanation may be that linear structures form during the collapse
of the protogalactic gas and when star formation turns on it triggers induced
star formation along the line of maximum density --- a process analogous to the
sequential propagation of OB star associations in molecular clouds in our own
galaxy.  However, there may be other mechanisms -- such as string-induced wakes
(Brandenberger 1991) --- which might more naturally produce a linear structure.

\section{Conclusion}

Using $HST$\ $I$-band images of a $K$-selected galaxy sample, we have shown
that the faint-end galaxy counts at $K \ge 19$\ are dominated by blue galaxies
with unusual morphologies and that at $K > 20$, the largest single class is a
new type of of galaxy with linear, often beaded, structure which we refer to as
a chain galaxy.

Based on indirect evidence and our limited spectroscopic information, we have
argued that the chain galaxies are lines of giant star-forming regions at
redshifts between 0.5 and 3.  If this interpretation is correct, then the
chains are very large galaxies in the process of formation.

The key question remaining is whether the chains are a transient population
of galaxies unrelated to present-day galaxies or whether they are their
lineal ancestors.  Songaila, Cowie \& Lilly (1990) have given
model-independent arguments that the blue galaxy population, of which the may
chains comprise a major part, do contain enough star formation to be the
progenitors of present-day galaxies.  However, in this case we would expect
to see `missing links' with morphologies intermediate between the chains and
current galaxies.  One candidate is the amorphous multicellular galaxies
described here and in Glazebrook \etal\ (1994b) which could represent the
first stage in the dissolution of the linear structure of the chains prior to
the onset of more normal spiral morphology.  Investigation of the kinematics
and colors of these objects should determine if this is a tenable hypothesis.

\acknowledgments

\newpage

\begin{figure}
\caption{(Plate~1) Cosmic-ray cleaned WFC2 images of the 13-hr field (SSA13)
taken in the F814W filter, and assembled as a composite of six 2600 s exposures
offset by 5 arcsec in successive exposures obtained over contiguous orbits.
Images from each of the three WFC chips (designated chips 2 through 4) are
shown, with chip 2 showing a stellar diffraction spike extending diagonally
across the frame.  The coordinate center for these observations is: $13^{\rm
hr}~10^{\rm m}~01^{\rm s}$, $43^{\circ}~00^{\prime}~32^{\prime\prime}$\ (1950),
and each chip is 80 arcsec on a side.  The final FWHM of the image is $0\p.19$
(no image restoration applied).  Each frame has N at the upper right corner and
E at the upper left.}
\end{figure}
%
\begin{figure}
\caption{(Plate~2) WFC2 images of the 22-hr field (SSA22) in the F814W filter.
As for Fig.~1, these images were assembled as composites of six
successive exposures of 2400 s each, obtained over contiguous orbits
on 21 November 1994, and similar set of six exposures obtained on 26
November 1994, for a total exposure of 8 hrs.  The final FWHM of the
image is $0\p.22$.  The field is centered on
$22^{\rm hr}~15^{\rm m}~01^{\rm s}$,
$00^{\circ}~00^{\prime}~00^{\prime\prime}$\ (1950).}
\end{figure}
%
\begin{figure}
\caption{A plot of the difference in $I_{HST}$ aperture magnitudes for
isolated objects computed with $1\p.6$ and $3\p.2$ diameters.  The spread in
total magnitudes corresponds to an uncertainty of approximately $\pm0.2$ mag.
Filled squares represent data from the SSA22 field; pluses are from SSA13.}
\end{figure}
%
\begin{figure}
\caption{Color conversion between $I_{HST}$ and the Kron-Cousins $I$
magnitudes of Paper I.  $(I_{HST}-I)$ is plotted vs \bmi\ color using the
measured $I$ and $B$ magnitudes from Papers I and II.  The least squares fit
gives a small color term: $-0.08+0.05$\bmi\ and a dispersion of 0.14 mag.
Squares are SSA22, pluses are SSA13.}
\end{figure}
%
\begin{figure}
\caption{$I$-band number counts from the combined SSA13 and SSA22 $HST$ fields.
The adopted magnitude limit of $I=26$ for the galaxy counts was determined from
tests on the incompleteness of object recovery at the faint end.  The dashed
histogram shows the raw $HST$ data points and the solid histogram the counts
corrected for incompleteness, with $1\sigma$ error bars indicated for each half
magnitude bin.  Filled squares show the points from the previous ground-based
work on the Hawaii Galaxy Survey fields (LCG), which extend to $I=24.5$.  The
thick solid lines show the $I$ counts from Tyson (1988).  The best power law
fits to ground-based and $HST$ $I$ counts are shown as the dotted line.}
\end{figure}
%
\begin{figure}
\caption{The $K$-band number counts of several groups are summarized here.
The histogram is the summary of Gardner \etal\ (1993) with dashed lines
indicating $1\sigma$\ error bars.  The closed box symbols are from Soifer
\etal\ 1994, the open boxes are points from McLeod \etal\ 1995, the filled
diamonds are from Djorgovski \etal\ 1995, and the stars are from the
present work. $1\sigma$ error bars are shown with each point.  The
agreement among the many works is remarkably good. The solid line shows the
power law fit given in the text.}
\end{figure}
%
\begin{figure}
\caption{The number counts in the $B$ (dotted), $I$ (solid), and $K$ (dashed)
color bands are compared.  $AB$ magnitudes have been used to reference all
these curves to a consistent zero-point.  The convergence of the $K$-band
number counts with the curves for the $I$- and $B$- bands at the faint
end is an indication of the strong blueing trend at faint magnitudes.}
\end{figure}
%
\begin{figure}
\caption{The \imk\ color distribution of a sample selected according to
ground-based $K$ magnitude.  At the brighter magnitudes ($K < 20$) the open
symbols show data from other fields in the Hawaii surveys.  At the fainter
magnitudes only the data from SSA13 (filled triangles) and SSA22 (filled
boxes) are shown using the $HST$\ $I$-band color.  The colors spread into
very red and very blue objects at faint $K$ magnitudes.  $1\sigma$ error
bars are displayed.  The dashed line shows the condition $(I - K) = 1.5$\ for
flat $f_{\nu}$\ galaxies.}
\end{figure}
%
\begin{figure}
\caption{The \bmi\ vs $I$ color distribution for objects with $I\leq25$,
where $I$ magnitudes have been determined from the full set of $HST$ WFC
data.  Data from SSA13 are shown as filled triangles, and from SSA22 as
filled boxes.  The dashed line shows the condition $(B - I) = 1.8$\ for
flat $f_{\nu}$\ galaxies.  The solid line is the $1~\sigma$\ upper limit on
\bmi .}
\end{figure}
%
\begin{figure}
\caption{(Plate~3) An atlas of all objects in the $HST$ WFC fields with
$K\leq18.5$ from the $K$-band strip measurements of Paper III.  Each object's
$K$ magnitude is shown at the lower left of the panel, and the redshift (or
designation, `star') is given at the lower right based on spectroscopic
identifications in Paper III, or subsequently taken with the LRIS spectrograph
on Keck (Songaila \etal\ 1995).  The gray scale has been set for each panel by
normalizing the image to the object's $I$-band flux, in order to provide
uniform contrast.  Consequently, sky noise levels appear enhanced for the
fainter objects. Each panel is $3\p.2$ arcsec on a side.}
\end{figure}
%
\begin{figure}
\caption{(Plate~4) An atlas of $18.5\leq K<19.5$ objects in the $HST$ WFC
fields.
Labeling, scaling, and field of view is given as for Fig.~10 (Plate 3);
$K$-band measurements come from the $K$ strips for the 22-hr field
(Paper~III), and the deep-field regions of the 13-hr field (Paper~I).}
\end{figure}
%
\begin{figure}
\caption{(Plate~5) An atlas of $19.5\leq K<20$ objects in the $HST$ WFC fields.
Labeling, scaling, and field of view is given as for Fig.~10 (Plate 3);
$K$-band measurements come from the deep-field regions of the 22-hr and
13-hr fields of Paper~I.}
\end{figure}
%
\begin{figure}
\caption{(Plate~6) An $I$-band atlas of a complete sample of $K\leq14$
objects from ground-based data.  The 25 spectroscopically identified
galaxies displayed here are shown as they would appear at $z = 0.3$\ with a
box size equal to that in Figs.~10--12.  The S/N is comparable to that of
the objects in Fig.~10, Plate 3.}
\end{figure}
%
\begin{figure}
\caption{(Plate~7) A $K=20.5\to22.0$ atlas of $HST$ images for objects in
the SSA22 deep field.  Field sizes and contrast levels are set following the
methodology used in Fig.~10.  Redshift information is sparse for this
faint-end sample, and the panels are labelled according to $K$ magnitude
and \bmi\ color.}
\end{figure}
%
\begin{figure}
\caption{(Plate~8) A $K=20.5\to21.5$ atlas of $HST$ images for objects in
the SSA13 deep field.  As for Fig.~14.}
\end{figure}
%
\begin{figure}
\caption{Distribution of galaxy types with $K$ magnitude according to the
rough classification scheme given in this paper: Type 1 having strong
nucleus and surrounding smooth envelope, Type 2 with strong nucleus and
apparent spiral structure, Type 3 amorphous, Type 4 merging or interacting,
and Type 5 chain galaxies.  Filled diamonds represent the magnitude range
$K\leq14$, filled squares $16\geq K\geq20$, and pluses $K>20$.  The
increase in amorphous (Type 3) and interacting (Type 4) objects can be
seen at the fainter $K$ magnitudes, with the majority of $K>20$ objects
falling into either amorphous (Type 3) or chain (Type 5) galaxies.}
\end{figure}
%
\begin{figure}
\caption{The fraction of Type 1 plus Type 2 galaxies as a function of
$K$ magnitude.  These classes correspond to the ellipticals, S0s and
spirals that constitute the bulk of `normal' galaxies in $K$-selected
samples.  As with Fig.~16, there is a noticeable decrease in such objects
at faint $K$ magnitudes.}
\end{figure}
\clearpage
\begin{figure}
\caption{(Plate~9) Atlas of $B<24.5$ selected $HST$ objects from the 13-hr
deep field.  Each frame is $3\p.2$ on a side, with the $B$ magnitude given
in the lower left corner and the redshift (where known) in the lower right
corner.  The very blue objects [\bmi$,1.4$] are labelled `flat' in the
upper right corner.}
\end{figure}
%
\begin{figure}
\caption{(Plate~10) Atlas of $B<24.5$ selected $HST$ objects from the 22-hr
deep field.  Labeling and field-of-view follows Fig.~16.}
\end{figure}
%
\begin{figure}
\caption{(Plates~11--13) An atlas of chain galaxies with $I < 25$\ in the
two fields.  Each image is $12\p.8$\ on a side.}
\end{figure}
%
\begin{figure}
\caption{The distribution of \bmi\ colors vs $I$ magnitudes for $HST$ WFC2
galaxies (pluses).  The chain galaxies are shown with filled squares.
Their onset at faint magnitudes and blue colors may be noted.}
\end{figure}
%
\begin{figure}
\caption{The profile along the longitudinal direction for a sample chain
galaxy is shown in solid lines, with the very narrow profile along the
transverse direction superposed (dashed lines) over one of the bright
`knots'.}
\end{figure}
%
\begin{figure}
\caption{Spectrum of SSA 22-16 plotted in the rest-wavelength based on a
redshift $z=1.36$. The hatched regions indicate the positions of strong
atmospheric absorption bands.  The redshift is based on identifying the
strong emission feature as \oii\ 3727.}
\end{figure}
%

{

\makeatletter
\def\jnl@aj{AJ}
\ifx\revtex@jnl\jnl@aj\let\tablebreak =\nl\fi
\makeatother

\begin{planotable}{rrrrrrr}
\tablewidth{42pc}
\tablecaption{Chain Galaxies}
\tablehead{
\colhead{} & \colhead{} &
\colhead{} & \colhead{} &
\colhead{} &
\colhead{Transverse $\sigma$} & \colhead{Axial~} \\[0.5ex]
\colhead{~~\#} & \colhead{~$I_{HST}$}  & \colhead{~~$K$} &
\colhead{$B$} & \colhead{\quad$(B-I_{HST})$\quad} &
\colhead{(arcsec)} & \colhead{Ratio}
}
\startdata
0 & 22.2 & 20.0 & 24.3 & 2.1 & 0.13 & 4.8 ~~\nl
1 & 22.5 & 21.1 & 24.4 & 1.9 & 0.14 & 2.2 ~~\nl
2 & 22.6 & \nodata & \nodata & \nodata & 0.11 & 3.4 ~~\nl
3 & 22.7 & 20.5 & 23.5 & 0.8 & 0.11 & 5.9 ~~\nl
4 & 22.9 & \nodata & \nodata & \nodata & 0.23 & 4.1 ~~\nl
5 & 23.2 & {\llap{$-$}}21.9 & 25.1 & 1.9 & 0.09 & 4.7 ~~\nl
6 & 23.3 & 20.9 & 24.1 & 0.8 & 0.13 & 4.3 ~~\nl
7 & 23.3 & 20.9 & 25.1 & 1.8 & 0.06 & 9.5 ~~\nl
8 & 23.4 & 20.5 & 24.3 & 0.9 & 0.05 & 8.0 ~~\nl
9 & 23.5 & 22.1 & 25.1 & 1.6 & 0.19 & 3.2 ~~\nl
10 & 23.6 & 23.4 & 24.8 & 1.2 & 0.12 & 3.8 ~~\nl
11 & 23.7 & {\llap{$-$}}21.4 & 25.2 & 1.5 & 0.11 & 3.7 ~~\nl
12 & 23.5 & 20.5 & 24.6 & 1.1 & 0.17 & 2.0 ~~\nl
13 & 24.0 & {\llap{$-$}}21.3 & 26.0 & 2.0 & 0.12 & 3.3 ~~\nl
14 & 23.9 & 21.7 & 25.5 & 1.6 & 0.10 & 3.9 ~~\nl
15 & 23.8 & 20.5 & 24.6 & 0.8 & \nodata & \nodata ~~\nl
16 & 24.0 & 21.3 & 25.9 & 1.9 & 0.06 & 8.5 ~~\nl
17 & 24.2 & 21.3 & 26.1 & 1.9 & 0.07 & 5.6 ~~\nl
18 & 23.3 & {\llap{$-$}}22.4 & 24.1 & 0.8 & 0.18 & 1.8 ~~\nl
19 & 24.5 & 21.8 & 26.3 & 1.8 & 0.06 & 5.3 ~~\nl
20 & 24.4 & 22.0 & 25.3 & 0.9 & 0.08 & \nodata ~~\nl
21 & 24.6 & 20.7 & 24.9 & 0.3 & \nodata & \nodata~~\nl
22 & 24.7 & 23.3 & 25.8 & 1.1 & 0.05 & 6.9 ~~\nl
23 & 24.9 & \nodata & \nodata & \nodata & \nodata & \nodata ~~\nl
24 & 25.0 & 21.6 & 27.4 & 2.4 & 0.10 & 5.7 ~~\nl
\noalign{\vskip -13pt}
\tablecomments{Where the enclosed total flux over the aperture is
negative, the magnitude shown reflects the absolute value of the
flux, with a leading minus sign.}
\end{planotable}
\clearpage
}

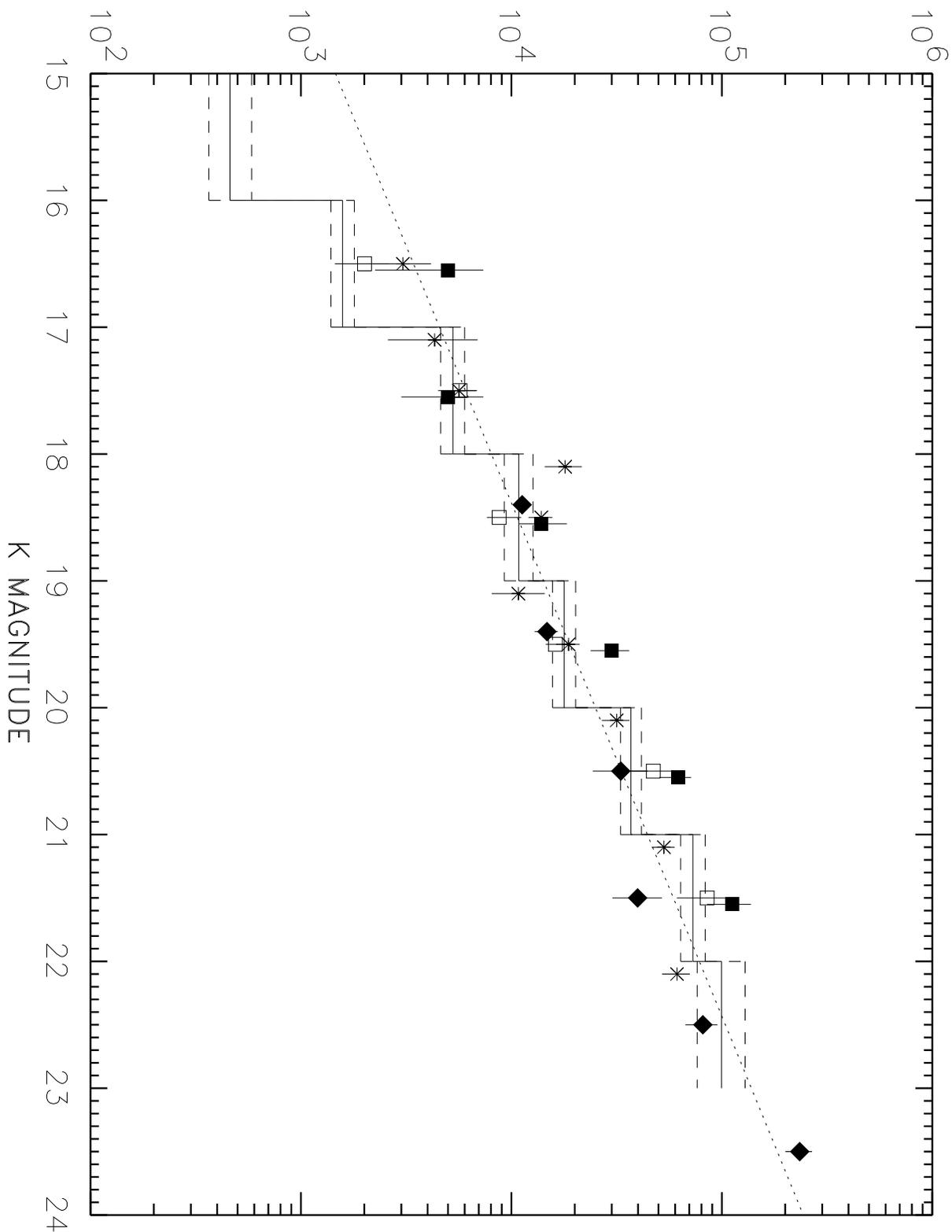
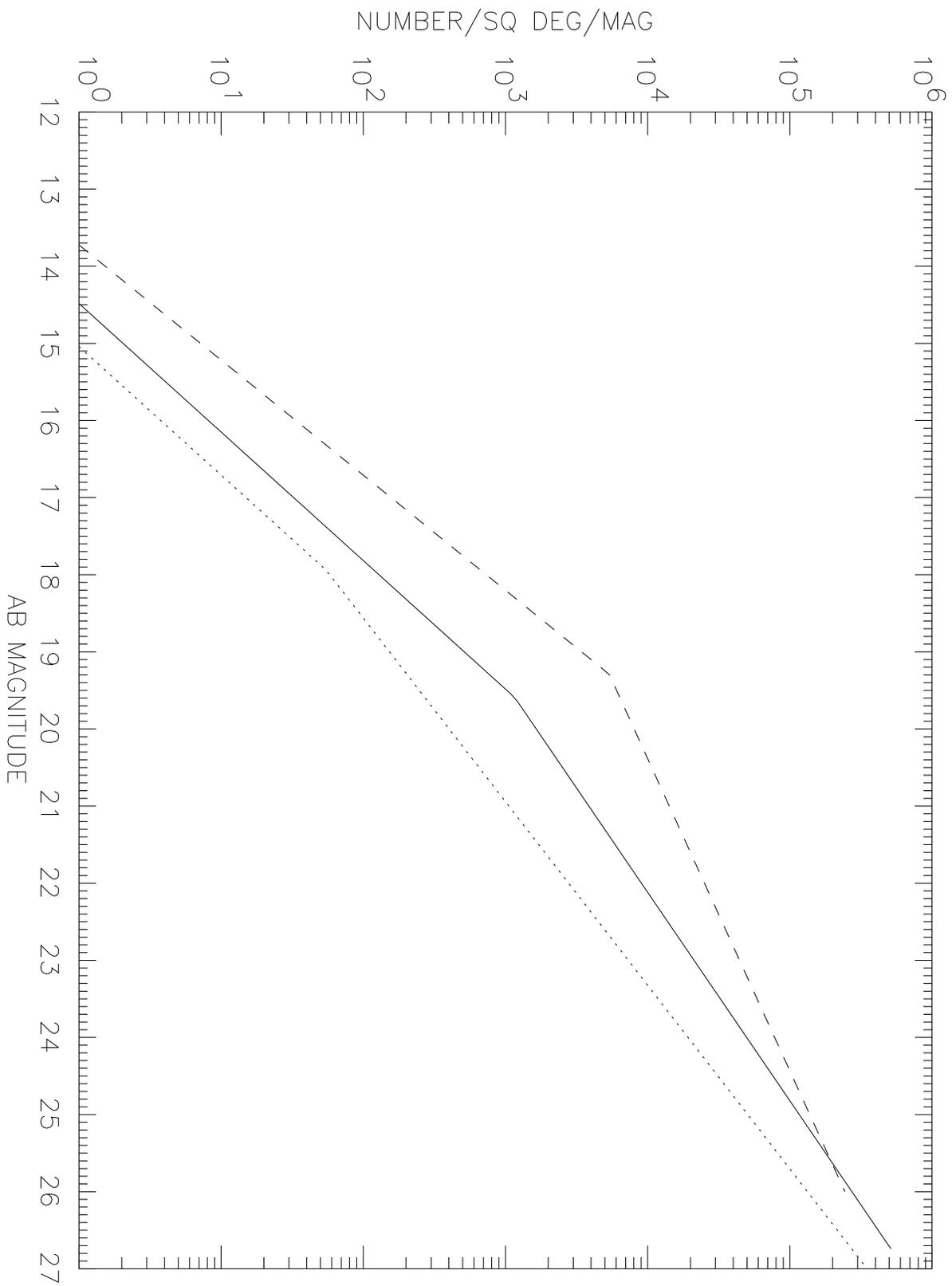
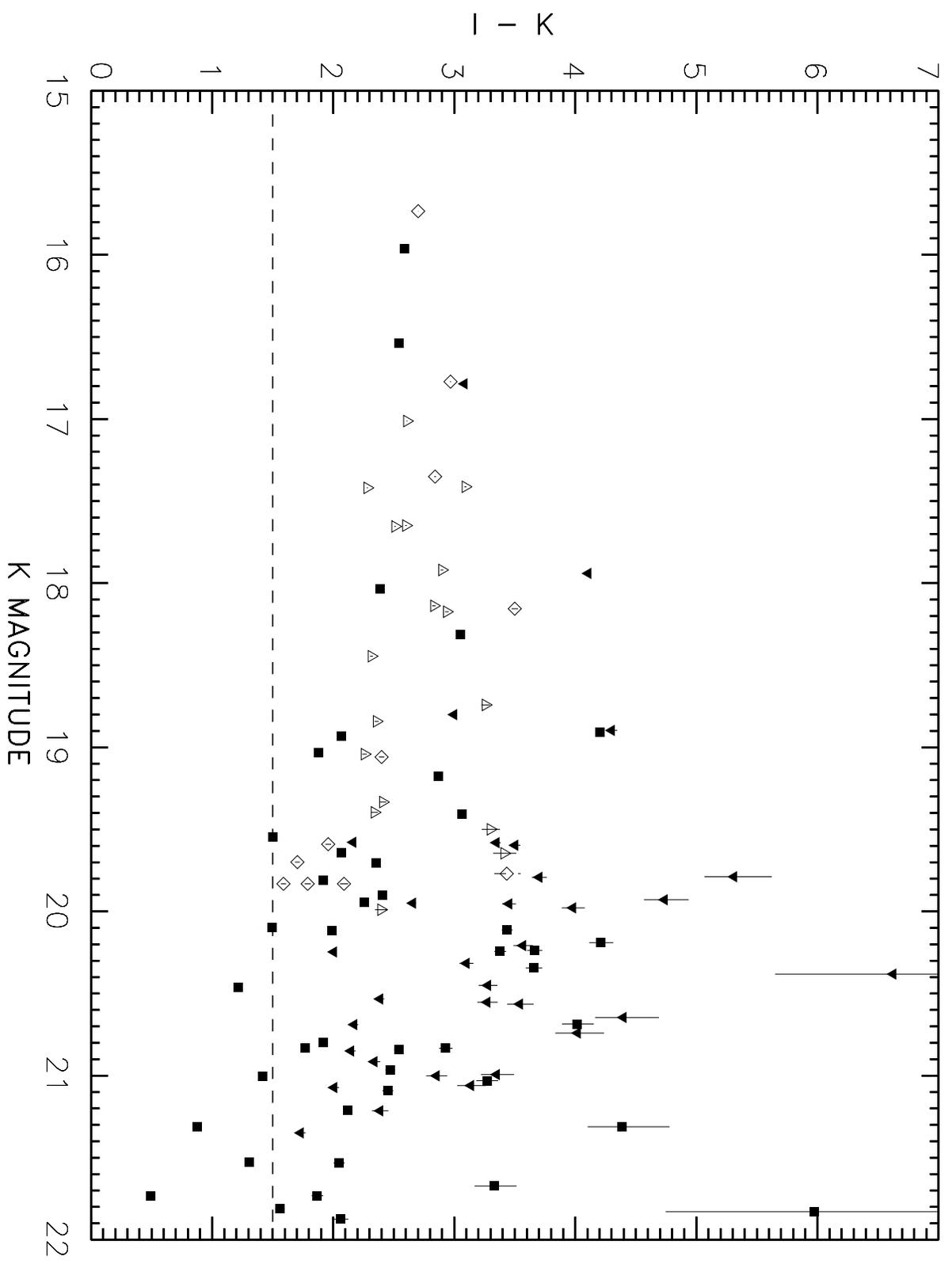
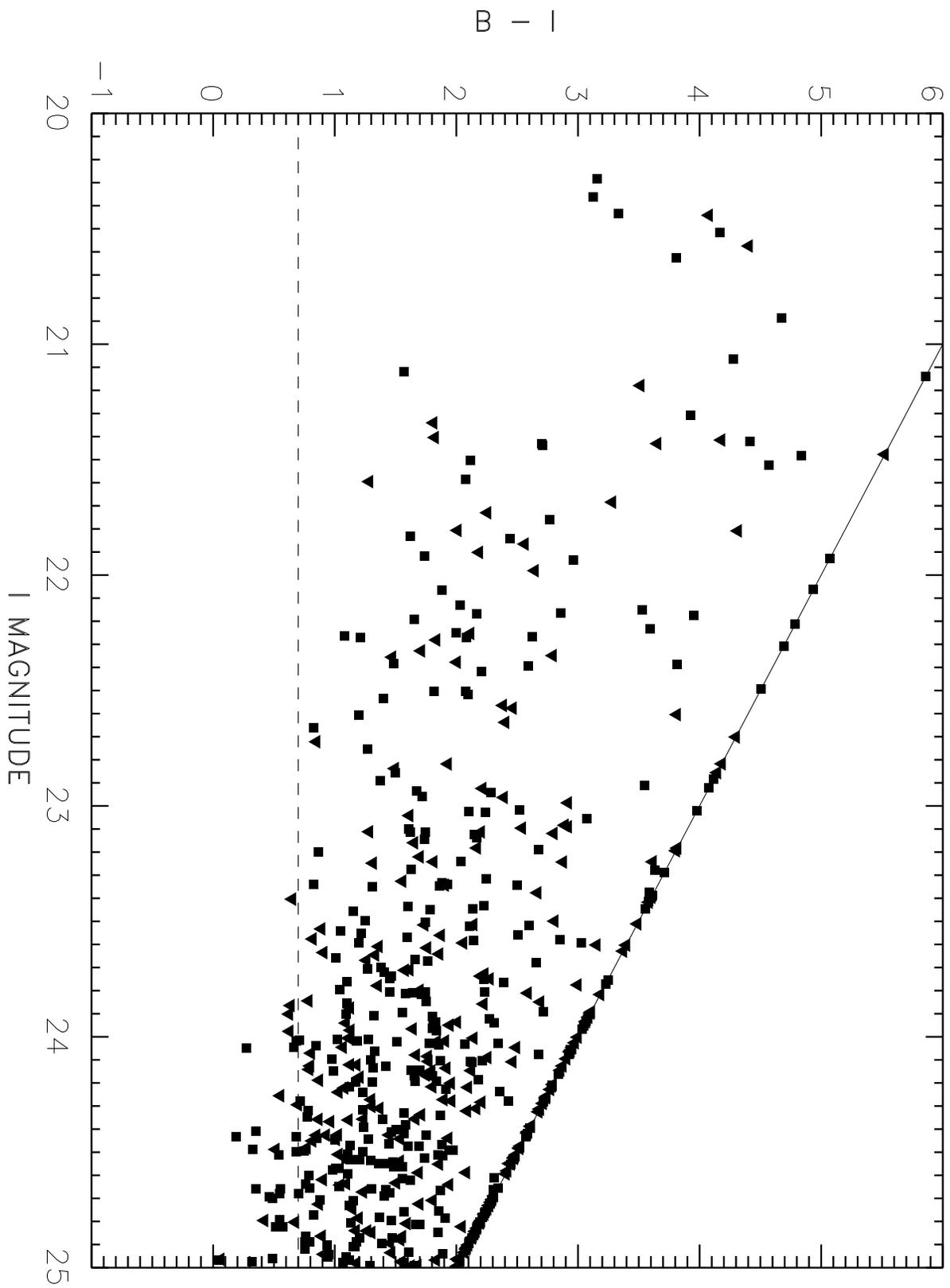
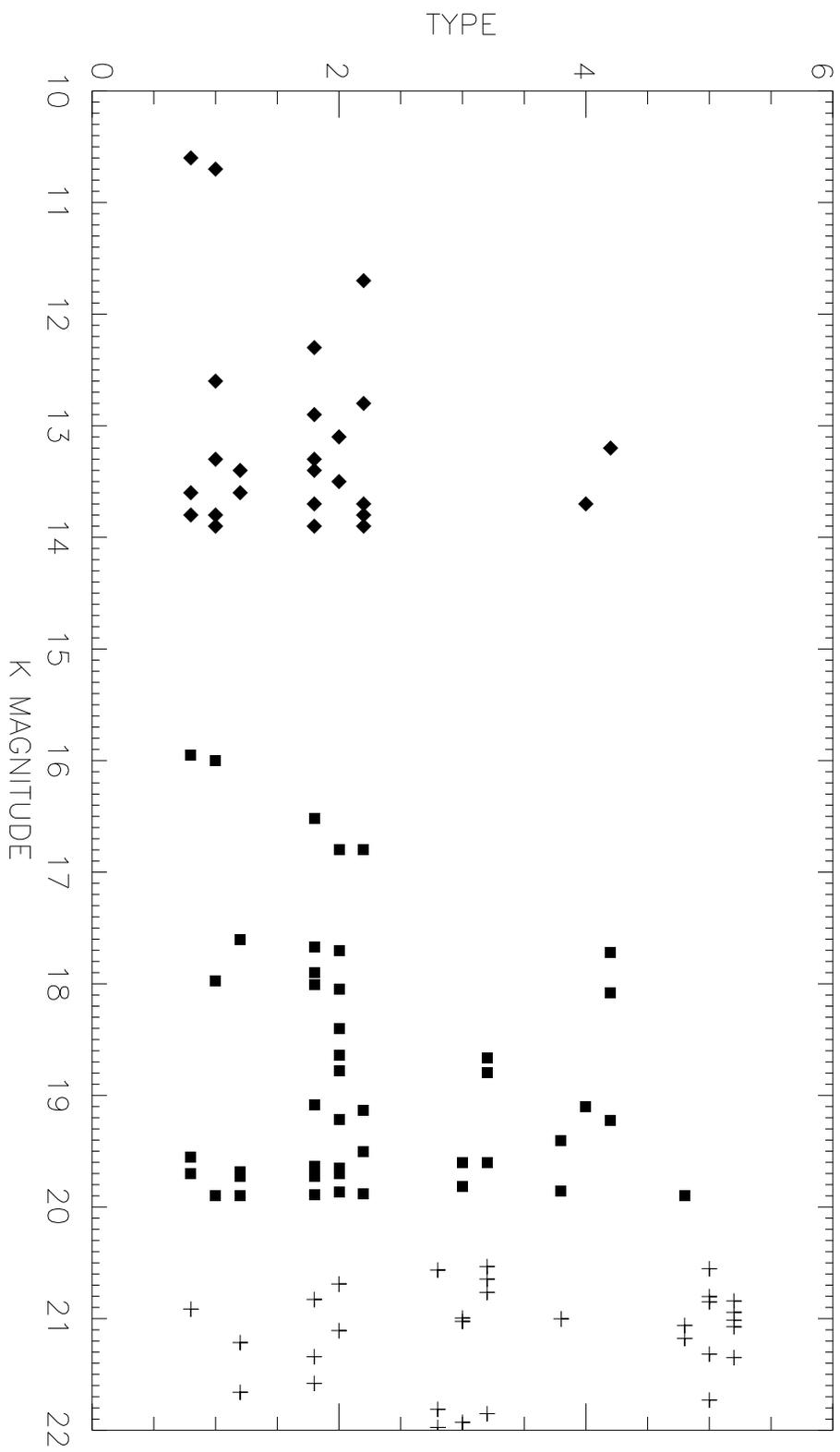
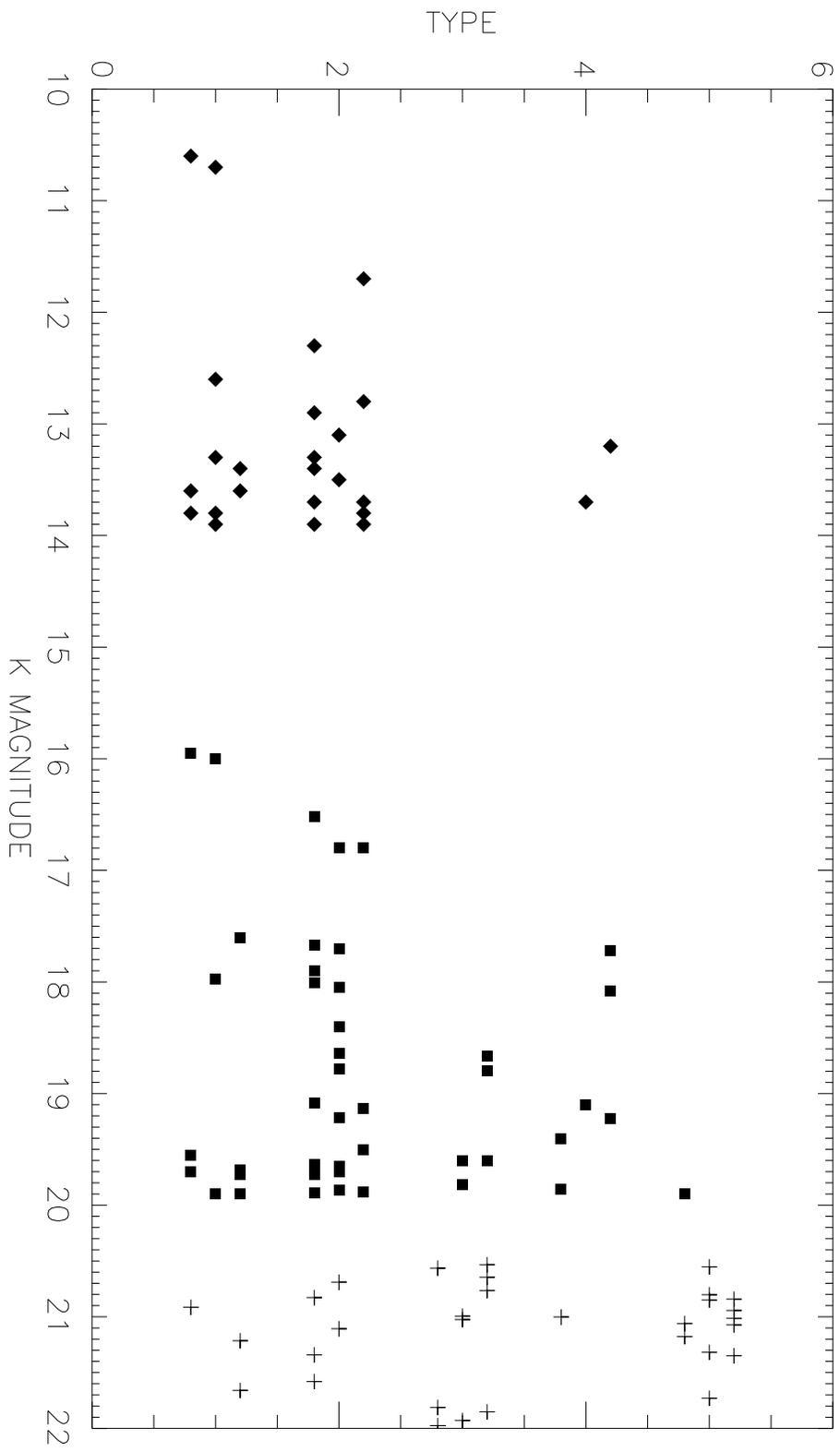
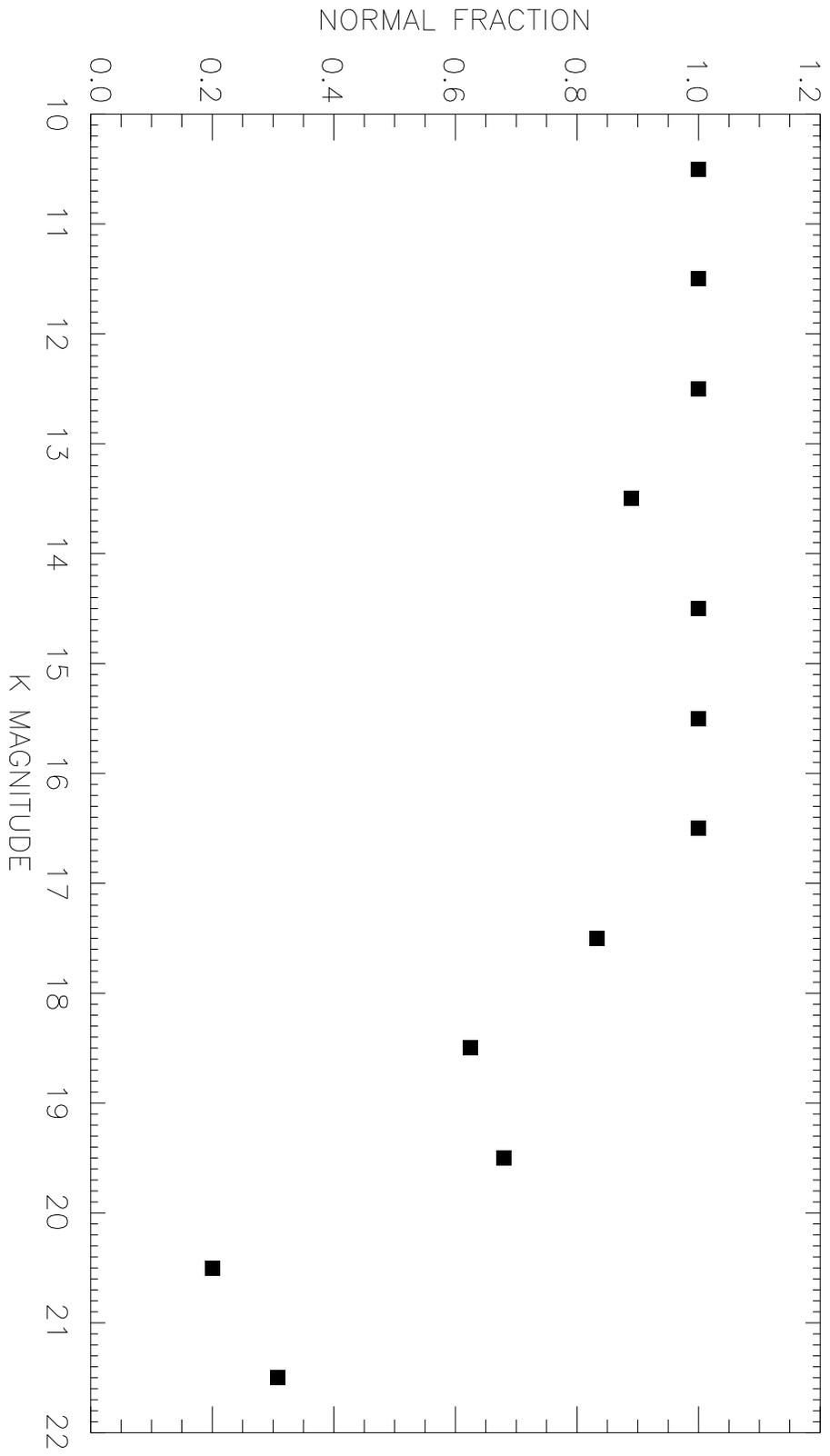
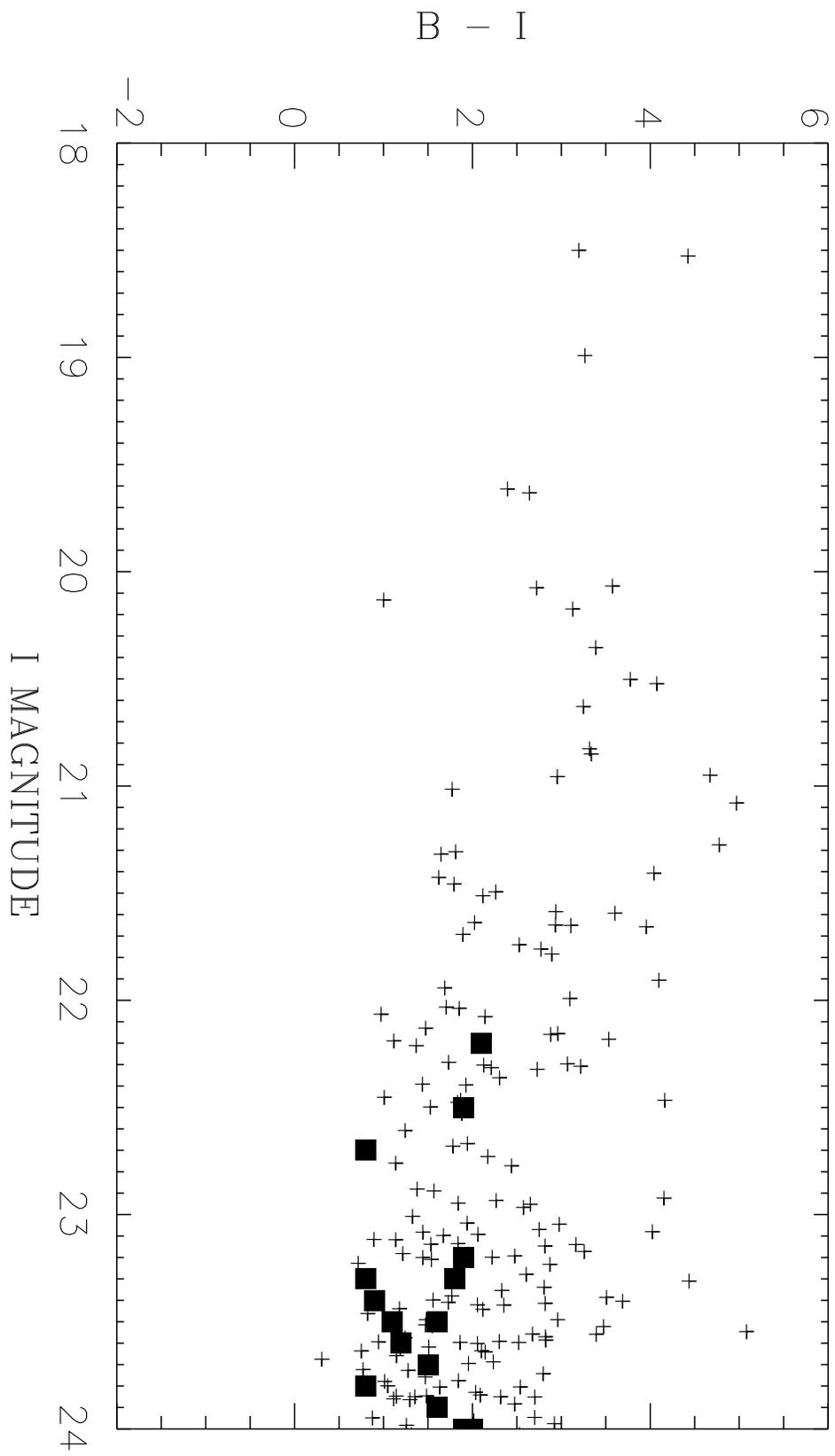
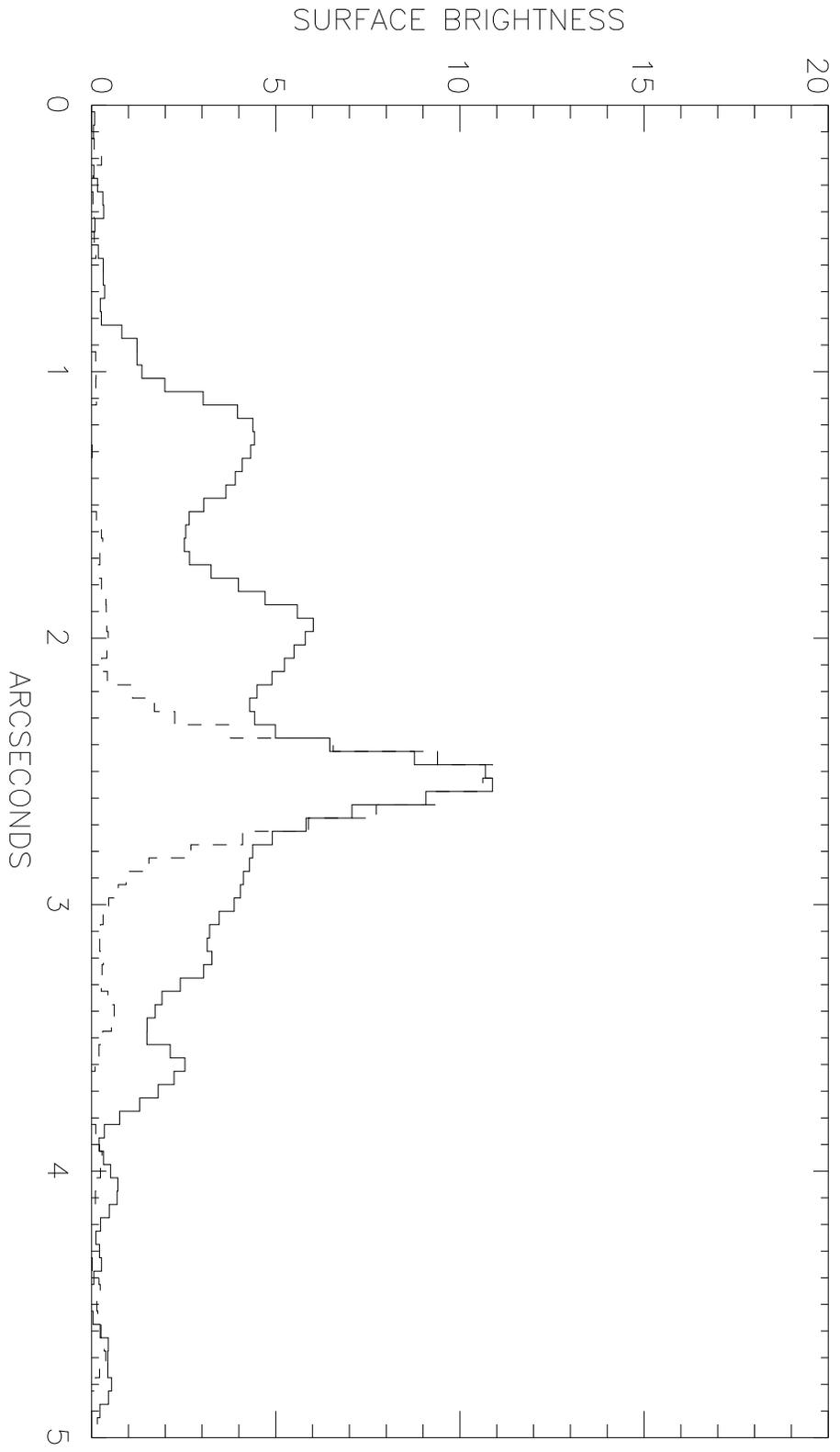
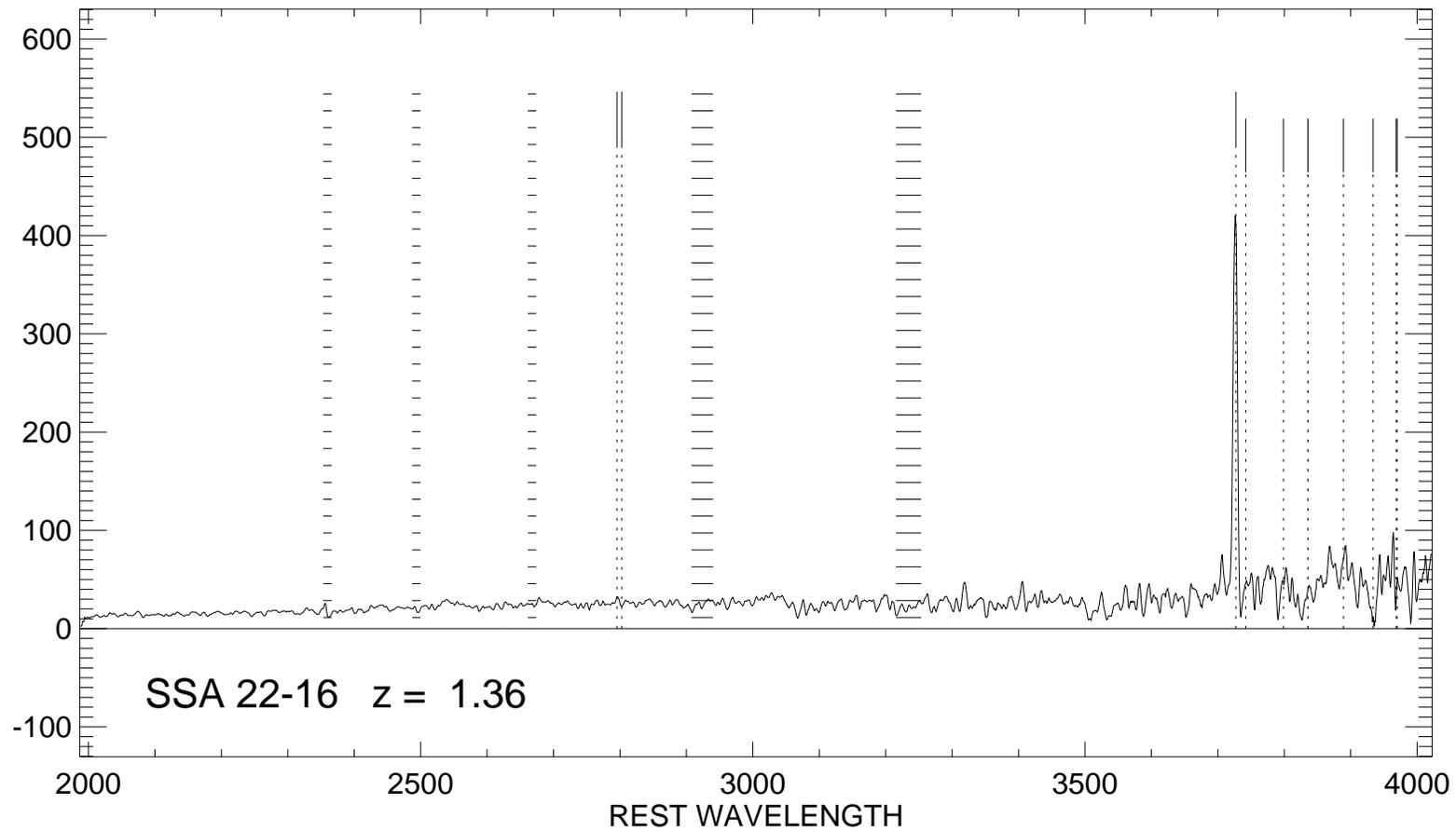
\end{document}